\documentclass[a4paper,usenatbib]{mnras}
\usepackage{newtxtext,newtxmath}
\usepackage[T1]{fontenc}
\usepackage{ae,aecompl}
\usepackage{graphicx,graphics}	
\usepackage{epsfig}
\usepackage{amsmath}				
\usepackage{amssymb}				
\newcommand{\pcm}{\,pc cm$^{-3}$}	

\title[Crab super-giant pulses]{Super-giant pulses from the Crab pulsar : Energy distribution and occurrence rate}
\author[A. Bera et al.]{
Apurba Bera$^{1}$\thanks{E-mail: apurba@ncra.tifr.res.in} and
Jayaram N. Chengalur$^{1}$\thanks{E-mail: chengalu@ncra.tifr.res.in} 
\\ \\
$^{1}$National Centre for Radio Astrophysics -- Tata Institute of Fundamental Research, Ganeshkhind, Pune - 411007, India	
}
\date{Accepted XXX. Received YYY; in original form ZZZ}
\pubyear{2019}

\begin{document}
\label{firstpage}
\pagerange{\pageref{firstpage}--\pageref{lastpage}}
\maketitle

\begin{abstract}

  We present statistical analysis of a fluence limited sample of over 1100 giant pulses from the Crab pulsar, with fluence $>$ 130 Jy~ms at $\sim$1330 MHz. These were detected in $\sim$260 hours of observation with the National Centre for Radio Astrophysics (NCRA)--15m radio telescope. We find that the pulse energy distribution follows a power law with index $\rm \alpha\approx-3$ at least up to a fluence of $\sim 5$ Jy s. The power law index agrees well with that found for lower energy pulses in the range 3--30 Jy ms. The fluence distribution of the Crab pulsar hence appears to follow a single power law over $\sim 3$ orders of magnitude in fluence. We do not see any evidence for the flattening at high fluences reported by earlier studies. We also find that at these fluence levels, the rate of giant-pulse emission varies by as much as a factor of $\sim$5 on time-scales of a few days, although the power law index of the pulse-energy distribution remains unchanged. The slope of the fluence distribution for Crab giant pulses is similar to that recently determined for the repeating FRB 121102. We also find an anti-correlation between the pulse fluence and the pulse width, so that more energetic pulses are preferentially shorter.
  
\end{abstract}



\section{Introduction}

The Crab pulsar is one of the handful pulsars known to emit giant radio pulses \citep[e.g.][]{ershov05aa, kazantsev17arep, mahajan18apj, mckee19mnras}. The giant pulses of the Crab pulsar reach flux densities several orders of magnitude higher than the average pulses \citep[e.g.][]{hankins03nature, jessner05asr, hankins07apj, crossley10apj}. Crab giant pulses are known to occur during both the pulsar's main-pulse and the inter-pulse emission window, although compared to the main-pulse, the frequency of occurrence is about an order of magnitude lower in the inter-pulse window \citep{karuppusamy10aa}. The relatively high rate of giant pulse emission makes the Crab pulsar best suited for statistical studies on giant pulses. In earlier studies, giant pulses have been found to show different statistical properties from those shown by the regular pulses. For example, the pulse-energy distribution of giant pulses follows a power law \citep[e.g.][]{popov07aa, bhat08apj, karuppusamy10aa}, while that of regular pulses follows exponential or log normal distribution \citep[e.g.][]{hesse74aa, ritchings76mnras},  This indicates that normal and giant pulses probably arise from different emission mechanisms.

In this letter, we present a statistical study of super-giant pulses (SGPs, fluence $\gtrsim$ 100 Jy ms) from the Crab pulsar detected at $\sim$ 1.4 GHz in $\sim$ 260 hours of observation, with pulse energies up to two orders of magnitudes above the energy range explored in similar studies in the literature. We compare the statistical properties of the super giant pulses we detect against those of their weaker cousins as well as the currently known properties of Fast Radio Bursts, for which one proposed model is super giant pulses from young pulsars or magnetars \citep[e.g.][]{lyutikov16mnras}.

\section{Observations \& data processing}

The Crab pulsar was observed for $\approx$ 260 hours in 31 observing sessions of typically 6--9 hour duration. The observations spanned over $\approx$ 45 days in February--April, 2019 and were carried out with the National Centre for Radio Astrophysics (NCRA)--15m telescope located in the NCRA campus in Pune, India. The observing bandwidth was 100 MHz between 1280-1380 MHz, (of which $\sim 65$~MHz is usable). Throughout the observing period, the telescope sensitivity (i.e. $\rm G/T_{sys}$) was measured using known bright radio sources - Cygnus-A, Cassiopeia-A and the Crab nebula. The sensitivity at 1330~MHz was measured to be $\rm G/T_{sys}\approx 3.4\times10^{-4}\;Jy^{-1}$ and was also constant to within 10\% over this period. 

\subsection{Real-time \& offline processing}

Real-time processing of the data was done using REDSPIDER-15m (REal-time De-disperser and Single Pulse IDentifiER) software developed by us for the NCRA-15m telescope. The software takes the Nyquist sampled data, computes the co- and cross-polar 128 channel spectra, which are averaged to a time resolution of $\approx 80 \mu$s. After RFI rejection to exclude narrow band as well as bursty broad band RFI, the data is incoherently de-dispersed. For our observing setup the intra-channel dispersion ($\sim$ 160 $\mu$s) is much larger than the intrinsic width of the Crab giant-pulse (i.e. $\sim$ few $\mu$s).  The incoherent de-dispersion was done at a fixed DM (i.e. that corresponding to the DM of the Crab pulsar, i.e. 56.77 \pcm \citep[e.g.][]{bilous16aa}. The two circular polarizations are averaged to produce the Stokes~I time series, which is searched for giant-pulse candidates, which are required to (1) have S/N $>$ 8, and (2) be brighter at the true DM than at DMs which are 5~\pcm above and below the true DM. An identical search is carried out on the time series after smoothing it by a boxcar kernel of width 2, with the same threshold for minimum S/N. (Recall that the observed pulse width is dominated by intra-channel dispersion which is $\sim 2$ times the time resolution.) The typical RMS noise of the system for time resolution of $\approx$ 80 $\mu$s is $\approx$ 54 Jy. The detection threshold (S/N $>$ 8) corresponds to a limiting fluence of $\approx$ 100 Jy ms for time resolution of $\approx$ 160 $\mu$s (after smoothing the time series). As the observed pulse width is dominated by intra-channel dispersion smearing, our detection threshold corresponds to a fluence and not a flux threshold. For all candidate giant-pulses, $\approx$ 150 ms (i.e. $\approx$ 3 times as long as the dispersion sweep time across the observation band for the DM of Crab) of Nyquist sampled raw data, centred at the pulse arrival time, is written to the disk for offline processing. All subsequent analysis uses the offline processed data. The data for each candidate pulse is coherently de-dispersed (yielding a time resolution of $\sim 80$~ns) and then visually examined to reject spurious pulses. About 20\% of the candidate pulses detected in the real-time pipeline are found to be spurious, most of which are close to the detection threshold.

\section{Results \& discussion}
\subsection{Pulse Shapes}

\begin{figure*}
\centering
\includegraphics[scale=0.48, trim={0.3cm, 0.5cm, 0.4cm, 0cm}, clip]{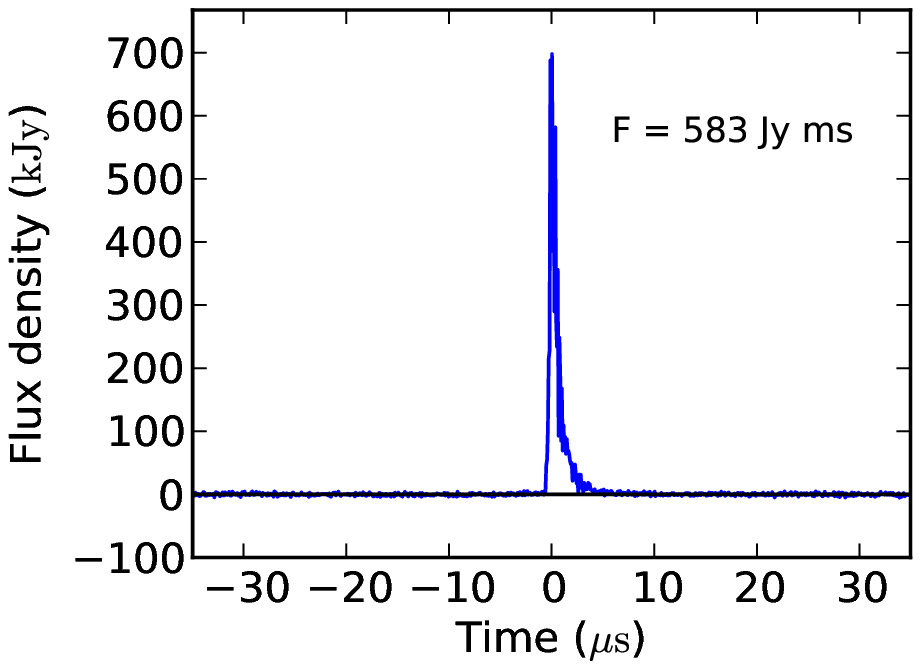}
\includegraphics[scale=0.48, trim={1.0cm, 0.5cm, 0.4cm, 0cm}, clip]{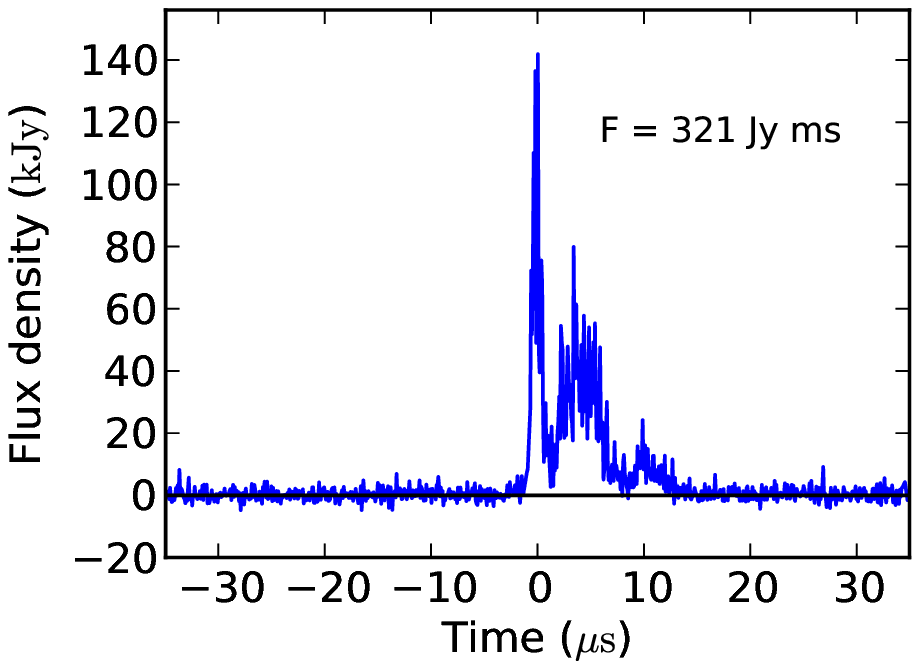}
\includegraphics[scale=0.48, trim={1.0cm, 0.5cm, 0.4cm, 0cm}, clip]{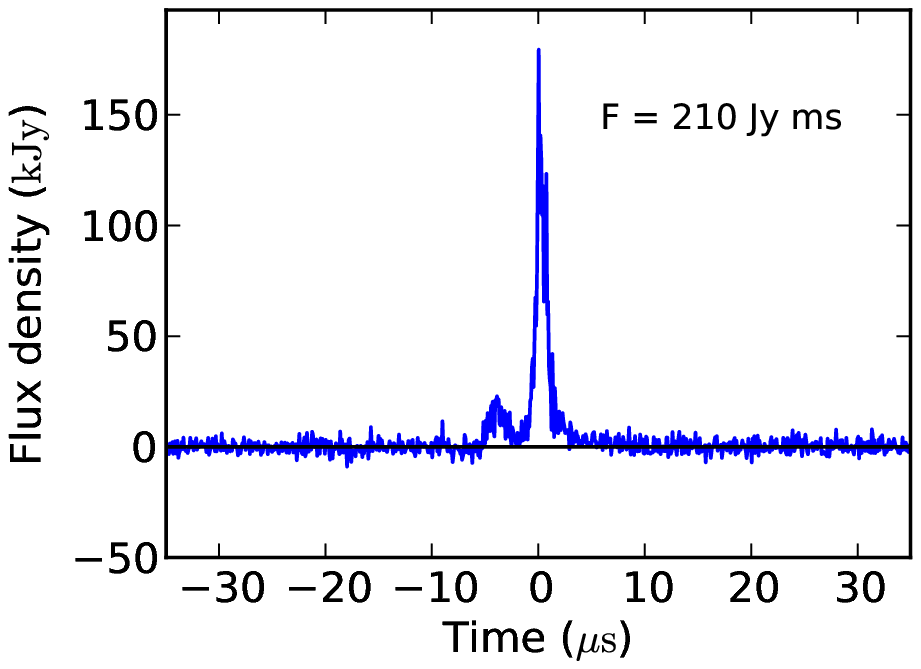}
\includegraphics[scale=0.48, trim={1.0cm, 0.5cm, 0.4cm, 0cm},clip]{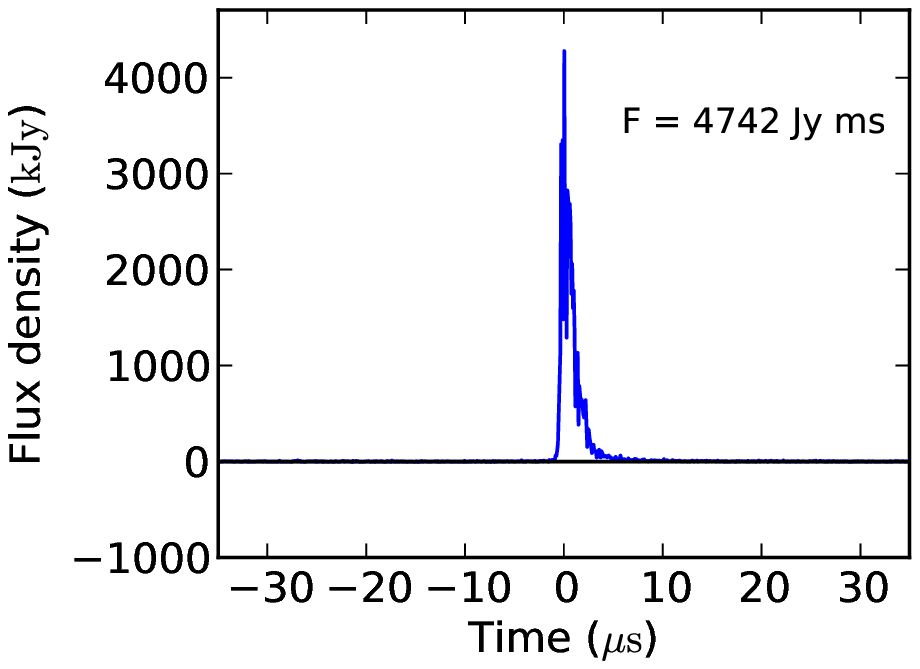}
\caption{Examples of super-giant pulses detected in our observation. From left to right -- [1] a typical single component pulse showing clear evidence of a scattering tail, [2] a multi component pulse with weaker trailing components, [3] a multi component pulse with a weaker leading component and [4] the most energetic super-giant pulse detected in the observation. The last one is also one of the most energetic giant pulses observed from the Crab pulsar at frequencies $\sim$1.4 GHz.}
\label{fig-pulses}
\end{figure*}

Our final sample consists of 1799 giant-pulses from the Crab pulsar detected in our observation. We have not aligned these pulses with respect to the average Crab pulse, and hence the pulses we detect correspond to giant pulses from both the main pulse and the inter-pulse. However, as noted above, earlier studies have shown that only $\sim 10$\% of the pulses correspond to the inter-pulse window. Our S/N threshold of 8 (at time resolution of 160 $\mu$s) corresponds to fluence threshold of $\gtrsim$ 100 Jy ms. However, for all of the following analysis, we conservatively include only pulses with fluence $\geq$ 130~Jy ms (corresponding to S/N~$\gtrsim$ 10) in order to avoid incompleteness effects  near the detection threshold. This yielded a complete sample of 1153 giant pulses with fluence $\geq$ 130~Jy ms.

Four example pulses are shown in Fig.~\ref{fig-pulses}. The detected pulses show a wide variety of temporal structures including single-component pulses and multi-component pulses with a strong burst preceded or followed by one or more weaker components. Each component shows narrow spiky features at much smaller time scales. The pulse seen in Fig.~\ref{fig-pulses}(a) shows a sharp rise followed by an slower decline, consistent with what would be expected from scattering. This is typical for all the single component pulses that we see. The pulse profiles that we see are broadly similar to those observed by \citet{sallmen99}. These authors discuss possible origins for the multiple components as well as the exponential like decay, concluding that the decay probably arises from scattering while the multiple components are likely intrinsic (and not copies caused by refraction/diffraction in discrete regions
of enhanced density). 

\subsection{Pulse-energy distribution}

\begin{figure}
\centering
\includegraphics[scale=0.7, trim={0.5cm 0.6cm 0.5cm 0.5cm}, clip]{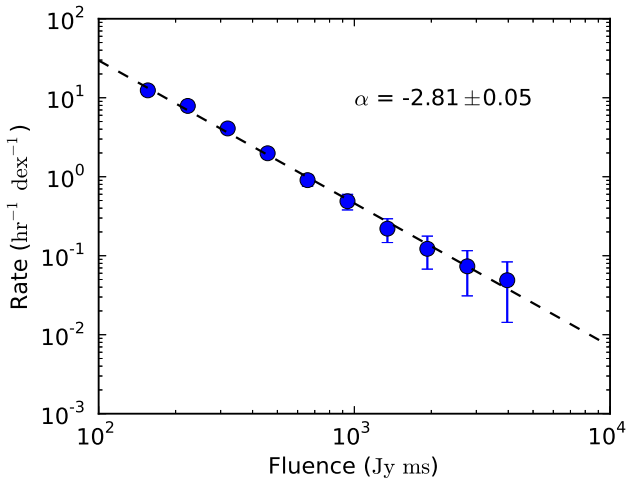}
\caption{Fluence distribution of the super-giant pulses from the Crab pulsar  for fluence F $>$ 130 Jy ms. The uncertainties associated with the data points correspond to Poisson errors. The dashed line shows a power law fit to the distribution with best fit power law exponent $\rm \alpha\approx-2.8$.}
\label{fig-fluence}
\end{figure}

We measure the fluence (which we use interchangeably with ``pulse-energy'') by integrating all the flux within a time interval of $\sim 3$ ms (i.e. $\approx$ the width of the average pulse profile) centred on each detected pulse.
The maximum separation of components in the multi-component pulses is $\sim$30~$\mu$s, well within our integration interval. The most energetic pulse in our sample has a fluence $\approx$ 4.7 Jy s and a peak flux density of $\approx$4 MJy. This is among the brightest known giant pulses of the Crab pulsar detected at $\sim$1.4 GHz. We parametrize the pulse energy distribution as

\begin{equation}
N(E)dE \propto E^{\alpha}dE
\end{equation}
where E is the fluence, $N(E)$ is the number of pulses per unit time per unit fluence. The fluence distribution, along with the best fit power law, which has a slope of  $\alpha=-2.81\pm0.05$ is shown in Fig.~\ref{fig-fluence}. The slope we find is within the range of  that observed earlier for giant pulses with much lower fluences \citep[e.g.][]{karuppusamy10aa, bhat08apj, popov07aa,mickaliger12, rudnitskii17}. \citet{cordes04} and \citet{mickaliger12} had seen signatures of flattening of the slope at high pulse energies, and had dubbed the giant pulses with fluxes above this flattening as ``super-giant pulses''. Both papers measure the pulse energies in terms of S/N ratios, which makes a direct comparison with our measurement difficult. However, the NCRA 15m with a 65~MHz usable bandwidth is about a factor of $\sim 8$ less sensitive than the system used by \cite{mickaliger12} (i.e. the NRAO~140ft with a 400~MHz usable bandwidth) even after accounting for the contribution of the Crab nebula to the system temperature (see e.g. the discussion in \cite{cordes04}). \cite{mickaliger12} see the flattening of the energy distribution at a S/N of $\sim 80$ (see their Fig. 6), which corresponds to about S/N $\approx 10$ for us, i.e. at fluences below what we are analyzing here.  Our large sample of bright pulses does not show any signature of this flattening.

\subsection{Pulse Fluence and Width}

\begin{figure}
\centering
\includegraphics[scale=0.7, trim={0.5cm 0.6cm 0.5cm 0.3cm}, clip]{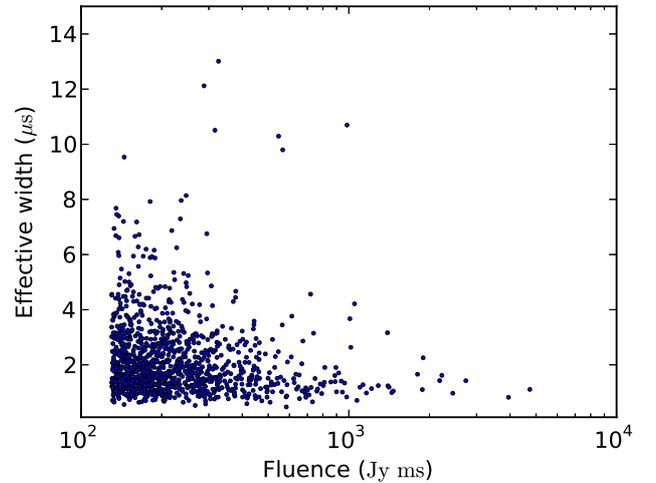}
\caption{Scatter plot of effective width (fluence/peak) and fluence of 1153 super-giant pulses with fluence$>$130 Jy ms. The effective width ranges from $\lesssim$1 $\mu$s to $\sim$ 10 $\mu$s, but most of the pulses have width between 1--2 $\mu$s.}
\label{fig-fluence_width}
\end{figure}  

\begin{figure*}
\centering
\includegraphics[scale=0.65, trim={0.4cm, 0.5cm, 0.5cm, 0cm}, clip]{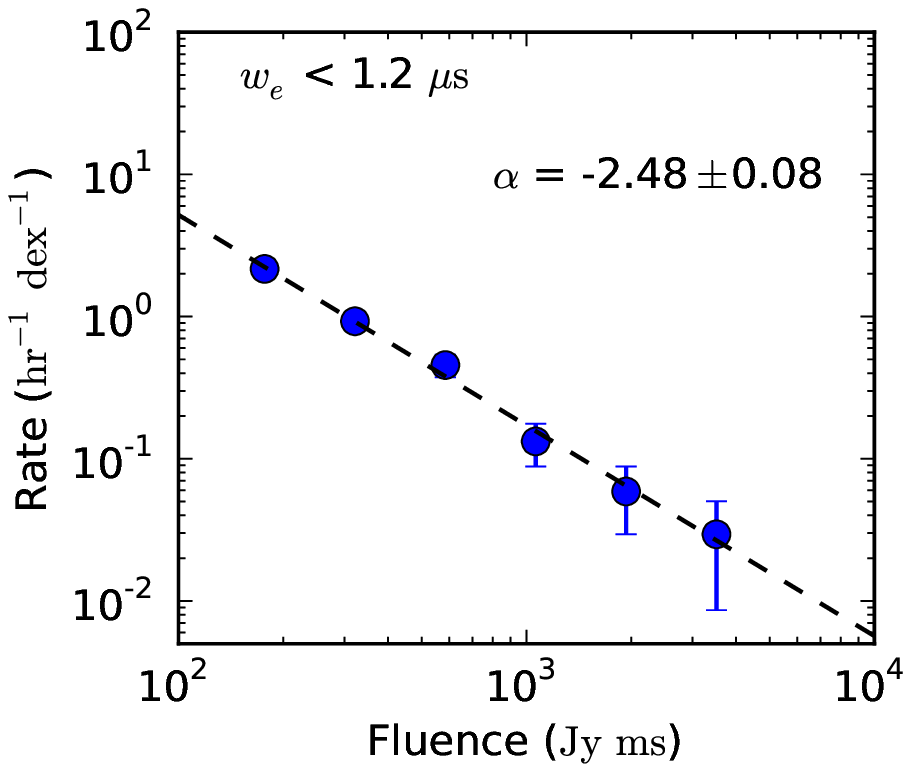}
\includegraphics[scale=0.65, trim={1.2cm, 0.5cm, 0.5cm, 0cm}, clip]{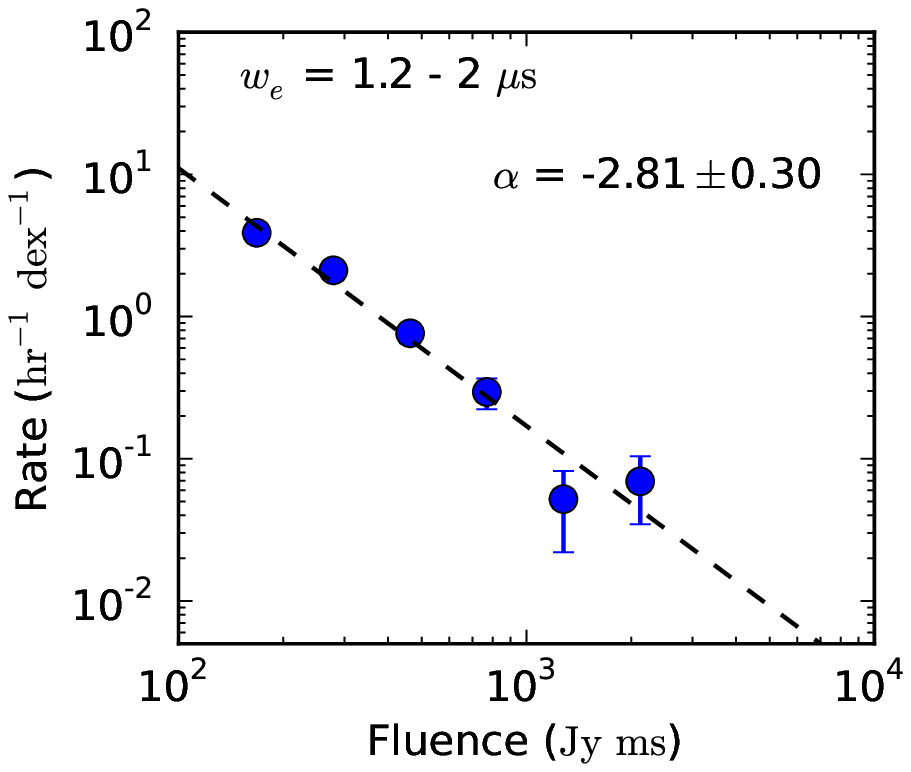}
\includegraphics[scale=0.65, trim={1.2cm, 0.5cm, 0.5cm, 0cm}, clip]{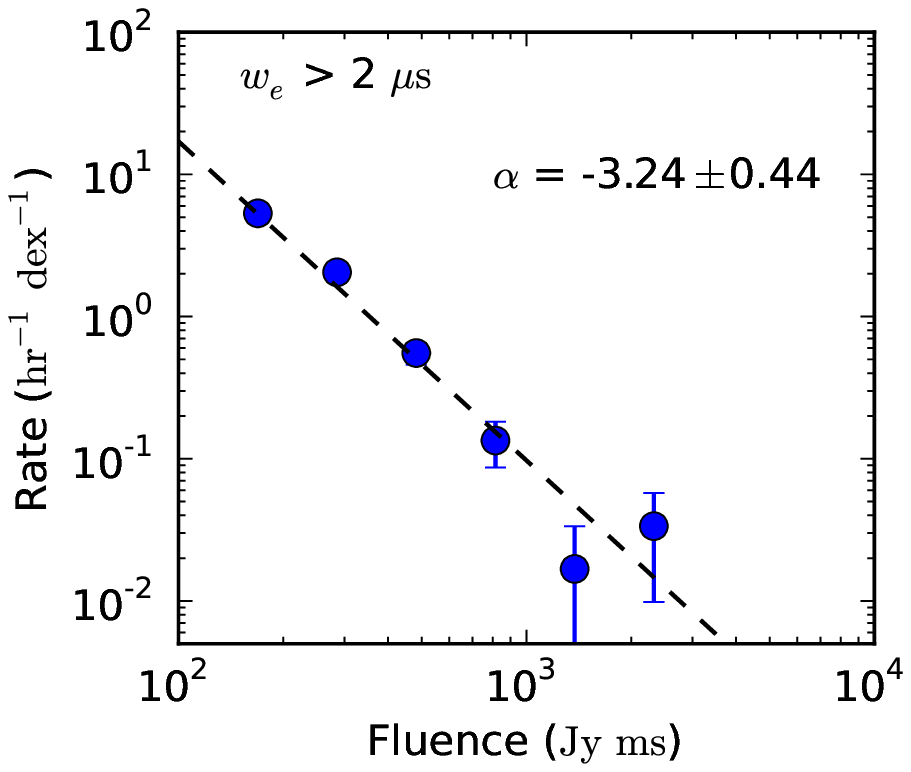}
\caption{Fluence distribution of the super-giant pulses (fluence F $>$ 130 Jy ms) for different effective widths, $<$1.2$\mu$s, 1.2--2$\mu$s and $>$2$\mu$s, respectively. The best fit power law index for each case is mentioned in the corresponding panel. The power law index is steeper for larger effective width, which implies that more energetic pulses are predominantly shorter. This trend is also apparent in figure \ref{fig-fluence_width}.}
\label{fig-fluence_distribution_width}
\end{figure*}

In situations where the pulse width is set entirely by scattering, one would expect that the pulse width would be independent of the fluence.  We define the ``effective width" ($w_e$) as the ratio of the fluence and peak flux density. Fig.~\ref{fig-fluence_width} shows a scatter plot of effective width and fluence. As can be seen, we see a clear anti-correlation, indicating that brighter pulses tend to be narrower. The widths we use average over the multiple components of the profile. \citet{sallmen99} used a parametric model to decompose complex profiles into multiple components. From the fit parameters to each component they find that the  widths did not correlate with the fluence, and argued that this supported the conclusion the pulse widths were dominated by scattering. We defer modelling of the profiles to a later paper, but our result here shows that over the pulse as a whole, the fluence anti-correlates with the width. \citet{popov07aa} and found a similar result for weaker giant pulses from the Crab pulsar. Given the large number of pulses that we detect, we can also look at the relationship between width and fluence is some detail. We show in Fig.~\ref{fig-fluence_distribution_width} the fluence distribution for different pulse widths. As can be seen, the pulse energy distribution is steeper for larger pulse widths, consistent with the anti-correlation between pulse width and fluence. This anti-correlation is in contrast with the situation for FRB pulses, where recent studies indicate that the pulse width appears to weakly correlate with the pulse luminosity \citep{hashimoto19}. Finally, we note that Fig.~\ref{fig-fluence} shows the typical effective width of our giant pulses is between 1--2 $\mu$s although there are pulses narrower than 1 $\mu$s and as wide as $\approx$10 $\mu$s. This is within the range of scatter broadening of the Crab giant pulses reported by \citet{rudnitskii17}.

\subsection{Occurrence rate }

\begin{figure}
\centering
\includegraphics[scale=0.65, trim={0.3cm 0.0cm 0.3cm 0.4cm}, clip]{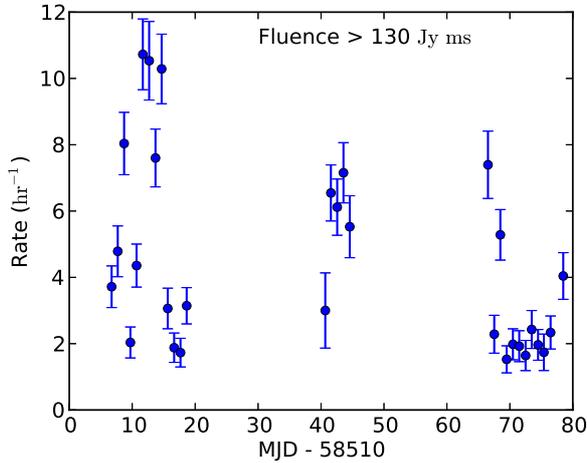}
\caption{Average rate of super-giant pulses (with fluence > 130 Jy ms) from the Crab pulsar in different observing sessions during February-April, 2019, as a function of the mean MJD of the observing sessions. Each observing session is typically 6-9 hours long. The uncertainties associated to data points correspond to Poisson errors. The blank days show when no observation was done.}
\label{fig-activity}
\end{figure}

\begin{figure}
\centering
\includegraphics[scale=0.7, trim={0.5cm 0.6cm 0.5cm 0.5cm}, clip]{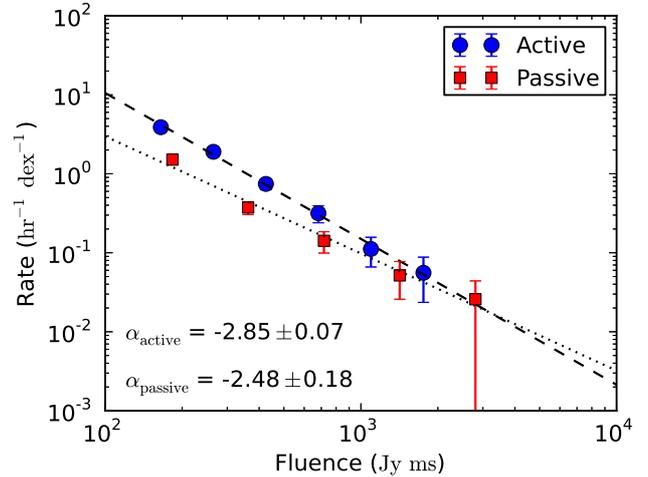}
\caption{Fluence distribution of the super-giant pulses (above fluence F $>$ 130 Jy ms) between 58521 $<$ MJD $<$ 58526 (``active" phase) and between 58579 $<$ MJD $<$ 58590 (``passive" phase). The uncertainties associated with the data points correspond to Poisson errors. The dashed lines show power law fits to the distributions. The best fit power law exponent remains same within the uncertainties in both the phases.}
\label{fig-fluence_activity}
\end{figure}

The giant pulse occurrence rate varies between the different observing sessions. Figure \ref{fig-activity} shows the average rate of occurrence of pulses (with fluence $>$ 130 Jy ms) in each observing session (of $\sim$ 6--9 hour duration) as a function of the mean MJD of the observing sessions. The average occurrence rate appears to vary on the time scale of a few days, by as much as a factor of $\sim$ 5. \citet{kazantsev19arxiv} also find a similar irregular variation in the rate of Crab giant pulses. In Fig.~\ref{fig-fluence_activity} we show the pulse-energy distribution separately for days when the occurrence rate was high (58521 $<$ MJD $<$ 58526) and when the occurrence rate was low (58579 $<$ MJD $<$ 58589). We find that although the occurrence rate has changed, the power law index is a constant within the uncertainties. This suggests that while the efficiency of the giant pulse emitting process appears to vary in time scale of a few days, intrinsic properties of the underlying process itself does not change. The intrinsic properties of the process may however evolve over longer time-scales; \cite{rudnitskii17} show variations in the slope of the pulse energy distribution for observations spread over a few years (albeit for pulses significantly weaker than the ones analyzed here).

\subsection{Crab giant pulses and FRBs}

The brightest pulse detected in our observation has a fluence of $\approx$ 4.7 Jy s. Modern FRB surveys have fluence limits of typically $\approx$ few Jy~ms (e.g. \cite{keane15}).  A pulse of fluence similar to the brightest pulse we have observed could only be detected to a distance of few 10s of kpc in these surveys. Models of FRBs which involve giant pulses from compact magnetized objects hence typically require sources with much higher magnetic fields, such as magnetars or extremely young pulsars (\cite{connor16,cordes16, katz16,lyutikov16mnras}). In this context, it is interesting that the the pulse-energy distribution of the repeating  FRB121102 \citep{spitler14apj, spitler16nature}, is well described by a power law with index $\alpha=-2.8\pm0.3$, (\citet{gourdji19apj}), in excellent agreement with what we find for the Crab giant pulses. Further, like the Crab giant pulses, the burst rate from FRB121102 also appears to vary with time \citep{oppermann18mnras, scholz16apj}. Although only a small fraction of the currently known FRBs have been known to repeat, the possibility that all FRBs repeat, cannot be ruled out with currently existing data \citep[e.g.][]{caleb19mnras}. However, the non detection of repeating bursts in long follow-up observations of some of the known non-repeating FRBs, with instruments sensitive enough to detect much fainter bursts than the original event, have put strong constraints on their repetition rates \citep[e.g.][]{petroff15mnras}. Reconciling these observations with giant pulse like sources would be easier if  the pulse-energy distribution were to flatten at high fluences. In 260 hours of observation of the Crab pulsar we see no evidence for any such flattening even at fluences as high as $\sim$ 5~Jy~s. However, if the repeating FRBs belong to a different class from the non-repeating ones, our results suggest super giant pulses emitted from highly energetic young pulsars or magnetars continue to be a viable model for repeating FRBs.

\section{Conclusions}

We studied the pulse-energy and occurrence rate statistics on a sample of 1153 giant pulses of the Crab pulsar, complete down to fluence $>$ 130~Jy ms, detected in $\sim$ 260 hours of observation at $\sim$ 1330 MHz with the NCRA 15m telescope. From these and earlier published results we find that the pulse-energy distribution of the giant pulses follows a power law with index $\alpha\sim-3$ from pulse-energies of $\sim$ 3 Jy ms all the way upto $\sim$ 5 Jy s, spanning three orders of magnitudes in fluence. We find no evidence for the earlier reported flattening of the distribution at high fluences. The power law index is steeper for wider pulses, implying that the more energetic pulses are preferentially narrower. We also observe that the average giant pulse occurrence rate varies by as much as a factor of $\sim$ 5 in time scale of a few days, although the power law index of the pulse energy distribution does not vary. This power law index is in good agreement with that recently determined for the repeating FRB~121102.

\section*{Acknowledgements}

We thank the staff members who have made these observations possible. The NCRA 15m telescope is run by the National Centre for Radio Astrophysics of the Tata Institute of Fundamental Research. AB thanks Avishek Basu for useful discussions.



\bibliographystyle{mnras}
\bibliography{frbrefs} 



%


\label{lastpage}
\end{document}